\documentclass[fleqn,twoside]{article}

\usepackage{amssymb}
\usepackage{times}
\usepackage{graphicx}
\usepackage{latexsym}
\usepackage{espcrc2}

\title{High Energy Tau Neutrinos}
\author{Jane H. MacGibbon\address{Code SN3, NASA Johnson Space Center, \\ 
		Houston Texas 77058 USA}
		and
         Ubi F. Wichoski \address{Depto. de F\'{\i}sica, CENTRA-IST, \\
                Av. Rovisco Pais, 1 - Lisbon 1049-001 Portugal}
	\thanks{Supported by ``Funda\c{c}\~ao para a Ci\^encia e a 
Tecnologia'' (FCT) under the program ``PRAXIS  XXI''.}}

\begin{document}

\begin{abstract}
The intrinsic tau neutrino flux from cosmological and astrophysical 
sources has usually been considered negligible in comparison to the 
electron and muon neutrino fluxes. However, the inclusion of the tau 
neutrino component coming from hadronic decay at the source can 
significantly modify the tau neutrino spectrum expected at Earth. 
We report our results on the high energy tau neutrino production and 
its implications for the observation of high energy neutrino events.
\end{abstract}

\maketitle

\section{INTRODUCTION}

Presently much effort is being expended to resolve the flavor of 
the incoming neutrino flux (for a review see \cite{Akhm}). One 
motivation for doing this is that the absolute sensitivity, or 
relative sensitivity compared with background, may be significantly 
greater for one neutrino species than for other species \cite{DRS}. 
Another primary motivation for identifying the neutrino flavor is to 
test the hypothesis that neutrinos undergo flavor oscillation as they 
propagate from source to Earth. Because the non-oscillation production 
of $\nu_{\tau}$ is 
assumed to be negligible relative to $\nu_{e}$ and $\nu_{\mu}$ production, 
and the cascade $\nu_{\tau}$ produced by collisions of primary $\nu_{e}$ and 
$\nu_{\mu}$ flux with the relic cosmic neutrino background is orders of 
magnitude less than the primary $\nu_{e}$ and $\nu_{\mu}$ flux, any 
detected $\nu_{\tau}$ component from a source outside the Solar System 
is currently expected to be indicative of oscillation. 
In this paper, we briefly report some results of our investigation into 
high energy $\nu_{\tau}$ production, presented elsewhere \cite{MWa}. 
We illustrate that the $\nu_{\tau}$ component at the high energy end 
in the hadronic decays is significantly higher than previously assumed 
and that this can have observational consequences.

\section{NEUTRINO PRODUCTION}

High energy neutrinos are expected to be produced by several 
cosmological and astrophysical sources, among others: active galactic 
nuclei (AGN) \cite{Ath}; topological defects (TD) such as 
superconducting, ordinary or VHS cosmic strings \cite{BS,HSW,WMB}; 
supermassive gauge and scalar particle (X-particle) decay or 
annihilation \cite{AHK}; and Hawking evaporation of 
primordial black holes (PBH) \cite{MC,MW,Hal}. Also, neutrinos can 
originate from the decay of photoproduced hadrons on the cosmic 
background radiation (CMB). 
In all these scenarios the production of tau neutrinos has been 
assumed to be negligible compared to the production of electron 
and muon neutrinos. 

We focus our attention on those cases where, irrespective of the scenario, 
there are sufficiently energetic interactions that quarks and gluons 
which fragment into jets of hadrons are expected 
to be produced. The spectra of these hadrons should be very similar 
to those measured in $e^+ e^- \rightarrow q \bar{q} \rightarrow$ 
\textit{hadrons} events in colliders. 
As a consequence, the final state particle distributions in these 
scenarios are expected to be dominated by hadronic decays at the source. 
In these decays over 90\% of the final state products will be 
$\pi^{0}$, $\pi^{+}$, and $\pi^{-}$, with the remainder mainly nucleons 
which decay into protons and antiprotons. On astrophysical timescales, 
the $\pi^{\pm}$ decay into $\stackrel{\mbox{\tiny (-)}}{\nu_{e}}$ and 
$\stackrel{\mbox{\tiny (-)}}{\nu_{\mu}}$ and $e^{\pm}$.  
The final cluster states (pions and nucleons) in QCD jets at 
accelerator energies can be loosely approximated by the Hill, 
Schramm, and Walker (HSW) fragmentation function 
\begin{equation}
\label{primdecay}
{{d N} \over {d x}} \, = \, {{15} \over {16}} x^{-3/2} (1 - x)^2 \, , 
\end{equation} 
where $x= E/m_J$ and $m_J$ is the total energy of the decaying jet 
\cite{HSW}. This distribution continues down to $E \sim 1$ GeV. When 
convolved with the $\pi^{\pm}$ decay, Eq. (\ref{primdecay}) leads to 
a similarly dominant $E^{-3/2}$ term in the $\nu_{e}$  and 
$\nu_{\mu}$ spectra at $x \lesssim 0.1$ and 
$d N_{\nu_{e},\nu_{\mu}}/dE \rightarrow 0$ as $x \rightarrow 1$. 
(For a full derivation of the $\nu_{e}$ and 
$\nu_{\mu}$ spectra using fragmentation function (\ref{primdecay}) 
see \cite{WMB}.) 

Because $\nu_{\tau}$ production is suppressed 
compared with other species, it has been assumed in previous 
astrophysical and cosmological flux calculations 
that the $\nu_{\tau}$ spectrum from hadronic jets is orders of magnitude 
less than the $\nu_{e}$ and $\nu_{\mu}$ spectra at all $x$. This is 
not so, once the energy of the jets surpasses the tau lepton and 
heavy quark masses. 
While indeed the total number of $\nu_{\tau}$ produced per jet is less than 
$10^{-3}$ of the total number of $\nu_{e}$ and $\nu_{\mu}$, the high 
$x$ tau neutrinos are predominantly produced by the initial decays of 
the heavier quarks with shorter lifetimes (the greatest contribution 
comes from the $t$ quark decay) and the $\nu_{e}$ and 
$\nu_{\mu}$ are produced by the final state cluster decays of the much 
lighter pions. This leads to significantly greater relative contribution 
from $\nu_{\tau}$ at high $x$ than previously assumed. The fragmentation 
distribution (\ref{primdecay}) is no longer relevant for the tau neutrino.

In Fig. 1 we show the $\nu_{\tau}$ spectrum, 
together with the $\nu_{e}$ and $\nu_{\mu}$ spectra, generated by 
the decay of 10 TeV $q \bar{q}$ jets. 
To simulate these spectra we used the QCD event generator HERWIG 
\cite{Cetal} and the process 
$e^+ e^- \rightarrow q_i \bar{q}_i \; (g)$, $i$ = all $q$ flavors. 
Consistent spectra are obtained with PYTHIA/JETSET (see \cite{MWa}). 
Note that in the region 
$0.1 \lesssim x \lesssim 1$, the tau neutrinos make up more than one 
tenth of the total neutrino contribution. Note also that below 
$x \lesssim 0.1$, $d N_{\nu_{\tau}}/d E$ falls off with roughly an 
$E^{-1/2}$ slope, and not the $E^{-3/2}$ slope of the 
$\nu_{e}$ and $\nu_{\mu}$ spectra.
\begin{figure}[htb]
\label{figure:nuspect}
\includegraphics[width=5.0cm,angle=-90]{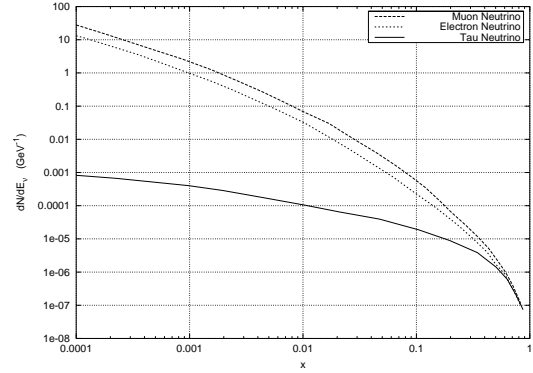}
\caption{The neutrino spectra, $d N/d E$ vs $x=E/m_J$, generated by 
the decay of $m_J=10$ TeV $q\bar{q}$ jets. The dashed line represents 
$\nu_{\mu}$, the dotted line represents $\nu_{e}$, and the solid line 
represents $\nu_{\tau}$.}
\end{figure}
From our 300 GeV \textendash { 75} TeV simulations, we find that 
the $\nu_{\tau}$  spectrum generated by $q \bar{q}$ jet decay can 
be parametrized as
\begin{eqnarray}
\label{fragtau}
{{d N_{\nu_{\tau}}} \over {d E_{\nu_{\tau}}}} &\simeq& 
\bigg(\frac{1}{2 m_J}\bigg) \Bigg[
0.15 \bigg(\frac{E_{\nu_{\tau}}}{m_J}\bigg)^{-1/2} - 
 0.36 + \nonumber \\
& & 0.27 \bigg(\frac{E_{\nu_{\tau}}}{m_J}\bigg)^{1/2} - 
0.06 \bigg(\frac{E_{\nu_{\tau}}}{m_J}\bigg)^{3/2} \Bigg]
\end{eqnarray}
per jet. 

Here we apply our results to two scenarios (for an expanded treatment see 
\cite{MWa}). The first one is the VHS cosmic string scenario. 
In the VHS scenario, 1 GeV \textendash { UHE} particle fluxes are 
generated by the decay of supermassive scalar and gauge particles 
emitted by the long strings over the age of the Universe. 
In Fig. 2 we show the neutrino spectra expected at Earth from VHS strings 
with a mass per unit length of $G \mu = 10^{-8}$ \cite{WMB}. 
The partial $\nu_{\tau}$ component which is solely produced in the 
collisions of the primary $\nu_{e}$ and $\nu_{\mu}$ with the relic 
cosmic $1.9^oK$ neutrino background \cite{Yosh} is also shown in 
Fig. 2. This collision-produced $\nu_{\tau}$ component 
is the only one presented in previous cosmic string papers. 
As can be seen, our results give a significant increase 
in the expected $\nu_{\tau}$ signal at the highest energies.   
\begin{figure}[htb]
\label{figure:fluxescasc}
\includegraphics*[width=5.0cm,angle=-90]{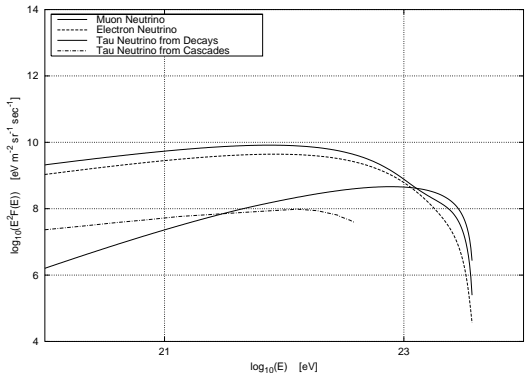} 
\caption{The neutrino spectra from VHS cosmic strings with a mass 
per unit length of $G \mu=10^{-8}$. The dash-dotted  curve is the 
$\nu_{\tau}$ component produced by cascades off the relic neutrino 
background \cite{Yosh} and the thick solid line is the the $\nu_{\tau}$ 
component produced by hadronic decays.} 
\end{figure}
In the second scenario, the evaporation of PBHs, we calculate 
the neutrino flux from the Hawking evaporation 
of a $T_{BH} \sim 100$ GeV PBH using HERWIG and following the method of 
Ref. \cite{MW}. 
The instantaneous emission per degree of freedom prior to decay is given by
\begin{equation}
\label{pbh}
\frac{d^{2} N_{\nu}}{d Q d t} = \frac{\Gamma_{s} ( T_{BH}, Q )}
{2 \pi \hslash} \bigg[e^{\frac{Q}{k T_{BH}}} - (- 1)^{2 s} 
\bigg]^{- 1} \;, 
\end{equation}
where $\Gamma_{s}$ is the absorption probability for 
species, $s$ is the spin, and $T_{BH}$ is the black hole 
temperature. The neutrino flux from the evaporating PBH including 
hadronic decays is shown in Fig. 3. 
\begin{figure}[htb]
\label{100GeVpbh}
\includegraphics[width=5.0cm,angle=-90]{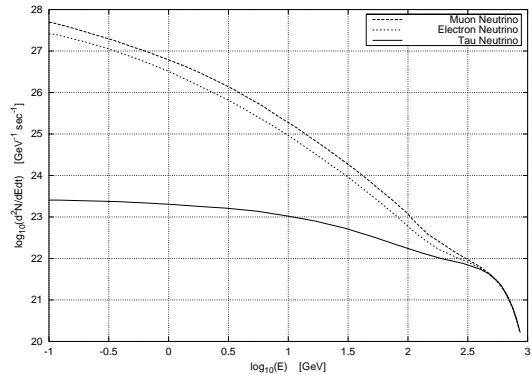} 
\caption{The neutrino flux from a $T_{BH} \sim 100$ GeV 
evaporating PBH at a redshift $(1 + z) \sim 1$ \cite{MWa}.The dashed 
line represents $\nu_{\mu}$, the dotted line represents $\nu_{e}$, and 
the solid line represents $\nu_{\tau}$.}
\end{figure}

\section{CONCLUSIONS}

Previous work has assumed  the $\nu_{\tau}$ component in TeV 
\textendash { UHE} hadronic decays to be negligible compared 
with the $\nu_{e}$ and $\nu_{\mu}$ components, at all neutrino energies. 
We find that this is not so once the decays are sufficiently 
energetic to include the heavier quarks. 
In particular, for neutrino energies 
in the decade below the energy of the initial decaying particle, the tau 
neutrino component is of similar magnitude to the $\nu_{e}$ and $\nu_{\mu}$ 
components. Below these energies the $\nu_{\tau}$ spectrum exhibits a 
slope of slightly less than $E^{-1/2}$  compared 
with the $E^{-3/2}$ slope of the $\nu_{e}$ and $\nu_{\mu}$ spectra. 
This analysis modifies the expected spectra in many astrophysical and 
cosmological high energy neutrino production scenarios. As a consequence, 
the observation of a significant $\nu_{\tau}$ to $\nu_{\mu}$ ratio 
at a given energy in high energy neutrino telescopes and detectors may 
be due to hadronic decay at the source and not 
$\nu_{\mu} \rightarrow \nu_{\tau}$ oscillation in transit. 
\\
Acknowledgments: 
It is a pleasure to thank M. Seymour and in particular B. Webber 
for advice, and R. Brandenberger and A. Heckler for encouragement.

\end{document}